\documentclass[12pt]{article}
\usepackage[margin=2cm]{geometry}
\usepackage{amsmath, amssymb, graphicx, tabularx, booktabs, amsthm, subfiles, enumitem, xcolor, braket, subcaption}

\numberwithin{equation}{section} 

\title{Rational Extension of Anisotropic Harmonic Oscillator Potentials in 
Higher Dimensions}

\author{
Rajesh Kumar$^{a,b}$\thanks{e-mail: kr.rajesh.phy@gmail.com (R.K.)}, 
Rajesh Kumar Yadav$^{b}$\thanks{e-mail: rajeshastrophysics@gmail.com (R.K.Y.)}, and 
Avinash Khare$^{c}$\thanks{e-mail: avinashkhare45@gmail.com (A.K.)}
}

\begin{document}

\maketitle

\begin{center}
\text{$^a$Department of Physics, Model College, Dumka-814101, India} \\
\text{$^b$Department of Physics, S. K. M. University, Dumka-814110, India} \\
\text{$^c$Department of Physics, Savitribai Phule Pune University, Pune-411007, India} \\
\end{center}

\begin{abstract}
This paper presents the first-order supersymmetric rational extension of 
the quantum anisotropic harmonic oscillator (QAHO) in multiple dimensions, 
including full-line, half-line, and their combinations. The exact solutions 
are in terms of the exceptional orthogonal polynomials. The rationally 
extended potentials are isospectral to the conventional QAHOs. 
\end{abstract}


\section{Introduction}
Supersymmetric Quantum Mechanics (SUSY QM) \cite{infeld1951factorization, gendenshtein1983derivation, cooper1983aspects, cooper1995supersymmetry, fellows2009factorization, mielnik1984factorization, hussin1998simple} is a powerful factorization method that has proven useful for generating new system of potentials from the known ones. In recent years, after the discovery of the exceptional orthogonal polynomials (EOPs) \cite{gomez2010extension, gomez2009extended}, a family of new potentials, isospectral to the corresponding conventional potentials were discovered \cite{quesne2008exceptional, odake2009infinitely, odake2010another, odake2013krein, yadav2013scattering, yadav2015scattering, yadav2015group, yadav2016parametric, grandati2012rational}. Apart from the other applications, rational extension of the isotropic harmonic oscillator was also done \cite{grandati2012rational}. The  extended eigenfunctions were expressed in terms of the exceptional Laguerre polynomials. Following this SUSY approach, a family of one-dimensional anharmonic oscillator potentials \cite{marquette2013two} which are strictly isospectral to the harmonic oscillator potential defined on the full-line have also been constructed for even co-dimension \(m\). Solutions of these potentials are obtained in terms of exceptional Hermite polynomials. In this case, it has been shown that the SUSY partner potential can have factorization energy above the ground state energy of the conventional potential. Building on this foundation and following the same approach, recently we constructed one-parameter family of rationally extended (RE) $m$-dependent potentials and studied some of the properties of these rationally extended potentials \cite{kumar2024rationally}.
While most of the studies in this area have so far concentrated on one-dimensional SUSY QM, rational extension of isotropic harmonic oscillator potentials in $D$-dimensions have been obtained using EOPs \cite{yadava2017rationally}. However, to the best of our knowledge, the rational extension of quantum anisotropic harmonic oscillators (QAHO) in higher dimensions has not been discussed in the literature so far. Moreover, relatively few efforts have been directed towards extending SUSY to higher dimensions \cite{clark1984non, de1983supersymmetric, andrianov1984factorization, khare1984supersymmetric, sukumar1985supersymmetry, leblanc1992extended, das1997higher}.
The purpose of this paper is to fill this gap by considering the rational extension of QAHO in two and higher dimensions. In particular by starting from a given QAHO in two and higher dimensions, we construct the higher dimensional rationally extended potentials using the SUSY approach. As discussed in \cite{marquette2013two}, the extended potentials defined on the full-line are restricted to even integers of \(m\) only. In order to include the odd \(m\) cases, we also consider the truncated QAHO on the half line and obtain the corresponding rationally extended potentials for all positive integer values of \(m\). We illustrate our approach by considering in detail the various possible combinations of half-line and full line QAHO and obtain the rational extension in all these cases.
In the two-dimensional case, we consider various combinations of full-line and half-line QAHOs and construct a family of corresponding RE two-dimensional harmonic oscillator potentials with their exact solutions in terms of exceptional Hermite and Laguerre polynomials. In the same way, we can generalize this to three or any higher dimensional QAHOs and solutions can be obtained easily in the Cartesian coordinates. We also consider a 3D QAHO and assume that the two of the three frequencies are equal (\(\omega_x=\omega_y\)) and obtain the RE potential and its solutions using the SUSY approach.

The paper is organized as follows: In Sec. 2, we provide a brief overview of the rational extension of the one-dimensional harmonic oscillator on the full line and then consider the truncated one dimensional harmonic oscillator on the half line and obtain the corresponding rationally extended potentials for any integral \(m\). In Sec. 3, we consider the various combinations of the two dimensional QAHO on the half-line and the full-line and in each case obtain the corresponding rationally extended potentials. The corresponding eigenfunctions are in terms of exceptional Hermite and Laguerre polynomials. In the same way one can generalize this to three and higher dimensions. As an illustration in Sec. 4, we extend the discussion to the QAHO in three dimensions. We discuss the case of QAHO with two out of three frequencies being equal and obtain the corresponding rationally extended potentials. Finally, in Section 5, we summarize our findings and suggest some open problems. In Appendix A and B we mention some of the well known results about SUSY in one and higher dimensions respectively which are being used in the present paper.


\section{One-Dimensional Harmonic Oscillator}
In this section, we first briefly review the results discussed in Refs. 
\cite{fellows2009factorization, marquette2013two} regarding the rational 
extension of a given one-dimensional potential \( V^+(x) \) defined as 
\begin{equation}\label{FLO:V+}
    V^+(x)=\frac{1}{4}\omega_x^2{x}^2,\qquad -\infty <x<\infty,
\end{equation}
where $\omega_x$ is the frequency of a particle moving along $x$-axis. Later we
consider the same one dimensional harmonic oscillator but on the half line and obtain its rational extension.
The eigenfunctions $\psi_n^+(x)$ and the energy eigenvalues $E^+_n$ of the potential 
(\ref{FLO:V+}) are well known  and are given by
\begin{equation}\label{FLO:psi+}
 \psi_n^+(x)\propto e^{-\frac{\omega_x}{4} x^2} H_n\left(\sqrt{\frac{\omega_x}{2}} x\right)\;  \nonumber\\	
\end{equation} 
and 
\begin{equation}
	E^+_n=\left(n+\frac{1}{2}\right)\omega_x\;,\quad n=0,1,2,\cdots
\end{equation}
respectively, where $H_n$ is the classical Hermite polynomial. The $m$-dependent
seedless function $\phi_m(x)$ is constructed by replacing $n\rightarrow m$ and 
$\omega_x\rightarrow -\omega_x$ in $\psi_n^+(x)$ given by
\begin{equation}\label{FLO:phi}
\phi_m(x) \propto\;e^{\frac{\omega_x}{4} x^2} 
H_m\left(i\sqrt{\frac{\omega_x}{2}} x\right)\,.
\end{equation}
Hence the partner potential $V^-(x,m)$ which is $m$ dependent is constructed 
using (\ref{FLO:phi}) and (\ref{F:W-1D}). Since the seedless function, 
$\phi_m(x)$, has the factorization energy 
\( \epsilon_m =-(m+\frac{1}{2})\omega_x\), which is less than the groundstate 
energy $E^+_0$ of $V^+(x)$ and therefore the $m$-dependent partner potentials 
$V^-(x,m)$ have an extra bound state with zero energy (for more details please 
see the appendix A). This potential is also known as the rational extension of 
the starting potential $V^+(x)$ defined for the even co-dimension of $m=0,2,4$ 
and so on. The form of this extended potential, its groundstate and the excited
state eigenfunctions are
\begin{align}
    \label{FLO:V-}
    V^-(x,m)&=V^+(x)-2\left[\frac{\mathcal{H}_m''\left(\sqrt{\frac{\omega_x}{2}} x\right)}{\mathcal{H}_m\left(\sqrt{\frac{\omega_x}{2}} x\right)}-\left[\frac{\mathcal{H}_m'\left(\sqrt{\frac{\omega_x}{2}} x\right)}{\mathcal{H}_m\left(\sqrt{\frac{\omega_x}{2}} x\right)}\right]^2+\frac{\omega_x}{2}\right], \quad -\infty<x<\infty\\
    \begin{split} \label{FLO:psi-}
        \psi^-_{0}(x,m)&=\phi_m^{-1}(x)\propto \frac{e^{- \frac{\omega_x }{4}x^2}}{\mathcal{H}_m\left(\sqrt{\frac{\omega_x}{2}} x\right)}\;,\qquad m=0,2,4,\cdots \\
       \mbox{and} \quad  \psi^-_{n+1}(x,m)&\propto\frac{e^{- \frac{\omega_x }{4}x^2}}{\mathcal{H}_m\left(\sqrt{\frac{\omega_x}{2}} x\right)} \hat{H}_{n+1,m}\left(\sqrt{\frac{\omega_x}{2}} x\right) \;,\qquad n=0,1,2,\cdots\\
    \end{split}
\end{align}
respectively, where, prime denotes derivatives with respect to $x$. The 
polynomial $\mathcal{H}_m(z)=(-i)^m H_m(i z)$ is the pseudo-Hermite polynomial 
and 
\begin{equation}
\hat{H}_{n+1,m}(z)=\left[\mathcal{H}_m(z)H_{n+1}(z) + H_{n}(z)\frac{d}{dz}\mathcal{H}_m(z)\right],
\end{equation}
is the Exceptional Hermite Polynomial with  \( n = -1, 0, 1, 2, \dots \) and 
\( \hat{H}_{0,m}(z) = 1 \). This system has co-dimension \( m \) and is 
orthogonal and complete with respect to the weight factor 
\( \frac{e^{-{z}^2}}{\mathcal{H}_m(z)^2} \). The energy eigenvalues of this 
extended potential using (\ref{F:psi-1D}) are given by
\begin{equation}\label{FLO:E-1D}
E^-_{n+1,m}=E^+_{n,m}=E^+_{n}-\epsilon_m=(n + m + 1)\omega_x \quad 
\text{ with }\quad E^-_{0,m}=0\,.
\end{equation}
It is worth repeating that in the expressions of \( V^-(x,m) \) the values of 
\( m \) are restricted to even integers only as the potential (\ref{FLO:V-}) 
diverges at the origin for odd integers \( m \). Instead,  
for odd \( m \), the potential \( V^-(x,m) \) is well defined on the half-line.
We now discuss this case in detail.

\subsection{Half-Line Oscillator}
Restricting to the positive real line, we define the half-oscillator potential 
$V_h^+(x)$ as
\begin{equation}\label{HLO:V+}
    V_h^+(x)=\begin{cases}
        \frac{1}{4}\omega_x^2 x^2 &,\qquad x>0\\
        \infty &,\qquad x\leq 0\,.
    \end{cases}
\end{equation}
It has two possible solutions distinguished by \(\alpha=\mp\frac{1}{2}\), out 
of which only one corresponding to \(\alpha=\frac{1}{2}\) is physically 
acceptable \cite{carballo2004polynomial,morales2015truncated} as it satisfies 
the right boundary condition at the origin. We however write both the solutions as they will be used to construct seedless functions for generating the RE potentials. The wavefunctions for this half-line oscillator potential in term 
of $\alpha$ and the classical Laguerre Polynomial 
$L_n^{(\alpha) }\left(\frac{\omega_x}{2} x^2\right)$ are given by
\begin{equation}\label{HLO:psi+}
\psi_{h,n}^+(x,\alpha)\propto x^{\alpha +\frac{1}{2}} 
e^{-\frac{\omega_x}{4} x^2} L_n^{(\alpha) }\left(\frac{\omega_x}{2} x^2\right);
\quad n=0,1,2,\cdots\,.
\end{equation} 
Similar to the full-line case, in this case one can easily construct the 
seedless function by transforming $n\rightarrow m$ and 
$\omega_x\rightarrow-\omega_x$ in $\psi_{h,n}^+(x,\alpha)$ as
\begin{equation}\label{HLO:phi}
\phi_{h,m}(x,\alpha)\propto x^{\alpha +\frac{1}{2}} e^{\frac{\omega_x}{4} x^2} 
L_m^{(\alpha )}\left(-\frac{\omega_x}{2} x^2\right),\qquad m=0,1,2,\cdots \,.
\end{equation}
The energy eigenvalues resulting from $\psi_{h,n}^+(x,\alpha)$ and 
$\phi_{h,m}(x,\alpha)$ with $V^+_h(x)$ potentials are
\begin{equation}
E^+_{h,n}(\alpha)=(2n+1+\alpha)\omega_x \qquad\text{ and}\qquad 
\epsilon_{h,m}(\alpha)=-(2m+1+\alpha)\omega_x
\end{equation}
respectively. Thus the energy eigenvalues for the Hamiltonian 
$H_h^+(x)$ is given similar to (\ref{E+(n,m)}) are
\begin{equation}\label{HLO:E+}
E^+_{h,n,m}(\alpha)=E^+_{h,n}(\alpha)-\epsilon_{h,m}(\alpha) = 
2(n+m+\alpha+1)\omega_x\,.
\end{equation}
Once one get the seedless function $\phi_{h,m}(x,\alpha)$, one can easily 
construct the RE potential $V^-_h(x,m,\alpha)$ using Eq. (\ref{F:V-1D}) for all positive integer values of $m$ and for $x \ge 0$. For $\alpha=\frac{1}{2}$, the RE potential is given by
\begin{align}\label{HLO:V-alpha+}
V_h^-(x, m, \frac{1}{2}) = V_h^+(x) - 2\left( \frac{\left[x L_m^{(\frac{1}{2})}\left(-\frac{\omega_x}{2} x^2\right)\right]''}{x L_m^{(\frac{1}{2})}\left(-\frac{\omega_x}{2} x^2\right)} - \left( \frac{\left[x L_m^{(\frac{1}{2})}\left(-\frac{\omega_x}{2} x^2\right)\right]'}{x L_m^{(\frac{1}{2})}\left(-\frac{\omega_x}{2} x^2\right)} \right)^2 + \frac{\omega_x}{2} \right); \qquad m = 0,1,2,\cdots \,.
\end{align}
It is worth pointing out that for $\alpha = 1/2$, Yadav et al. 
\cite{yadava2017rationally} had 
already obtained similar potential in arbitrary $D$ dimensions and our result is a special case of theirs in case \( D = 1, \; l = 2\), and 
$\omega_x = \omega$. 
However the potential \(V_h^-(x,m,\alpha)\) for $\alpha=-\frac{1}{2}$ is new and is given by 
\begin{align}
\begin{split}\label{HLO:V-alpha-}
V_h^-(x,m,-\frac{1}{2}) &= V_h^+(x) - 2\left( \frac{L_m^{(-\frac{1}{2})''}
\left(-\frac{\omega_x}{2} x^2\right)}{L_m^{(-\frac{1}{2})}
\left(-\frac{\omega_x}{2} x^2\right)} - \left( \frac{L_m^{(-\frac{1}{2})'}
\left(-\frac{\omega_x}{2} x^2\right)}{L_m^{(-\frac{1}{2})}
\left(-\frac{\omega_x}{2} x^2\right)} \right)^2 + \frac{\omega_x}{2} \right);
\qquad m=0,1,2,\cdots \,.
    \end{split}
\end{align}
In Table $1$, we have given expressions for $V_{h}^{-}(x,m,\alpha = 1/2)$ and
$V_{h}^{-}(x,m,\alpha = -1/2)$ in case $m = 0,1,2,3$. 
The wavefunctions of both systems ($\alpha=\pm\frac{1}{2}$) are obtained using 
(\ref{F:psi-1D}) and (\ref{F:A-1D}) and are given by 
\begin{equation}\label{HLO:psi-}
\psi^-_{h,n}(x,m,\alpha)\propto x^{\alpha +\frac{3}{2}} e^{-\frac{\omega_x}{4}x^2}\frac{\hat{L}_{n,m}^{(\alpha)}(\frac{\omega_x}{2}x^2)}{L_m^{(\alpha) }\left(-\frac{\omega_x}{2}x^2  \right)}\,
\end{equation}
where, \( \hat{L}_{m}^{(\alpha)}(s) \) is the exceptional Laguerre polynomial 
given by
\begin{equation}
 \hat{L}_{n,m}^{(\alpha)}(s)=L_{m}^{(\alpha)}(-s) L_n^{(\alpha+1)}(s)+L_{m-1}^{(\alpha+1)}(-s) L_{n}^{(\alpha)}(s)\,.
\end{equation}
In Table $2$, we have given expressions for $\psi_{h,n}^{-}(x,m,\alpha = 1/2)$ and $\psi_{h,n}^{-}(x,m,\alpha = -1/2)$ in case $m = 0,1,2,3$. It is worth 
pointing out that whereas for $\alpha = 1/2$ case, all the eigenfunctions 
for small $x$ behave like $x^2$, for $\alpha = -1/2$ case, all the 
eigenfunctions for small $x$ behave like $x$. However the asymptotic behavior of all the eigenfunctions in both the cases is the same. In Figure 1, we have plotted the eigenfunctions $\psi_{h,n}(x,m,\alpha=\pm 1/2)$ as a 
function of $x$ in case $m = 1$ and $n = 0,1,2,3$.  
The energy eigenvalues corresponding to the Hamiltonian $H_h^-(x)$ for the 
potential $V_h^-(x,m,\alpha)$ are strictly isospectral to $H_h^+(x)$ and are given by
\begin{equation}\label{HLO:E-}
E^-_{h,n,m}(\alpha)=E^+_{h,n,m}(\alpha)=2\left(n+m+1+\alpha\right)\omega_x\,.
\end{equation}

\begin{figure}[htp]
    \centering
    \begin{subfigure}[b]{0.48\textwidth}
        \centering
        \includegraphics[width=\textwidth]{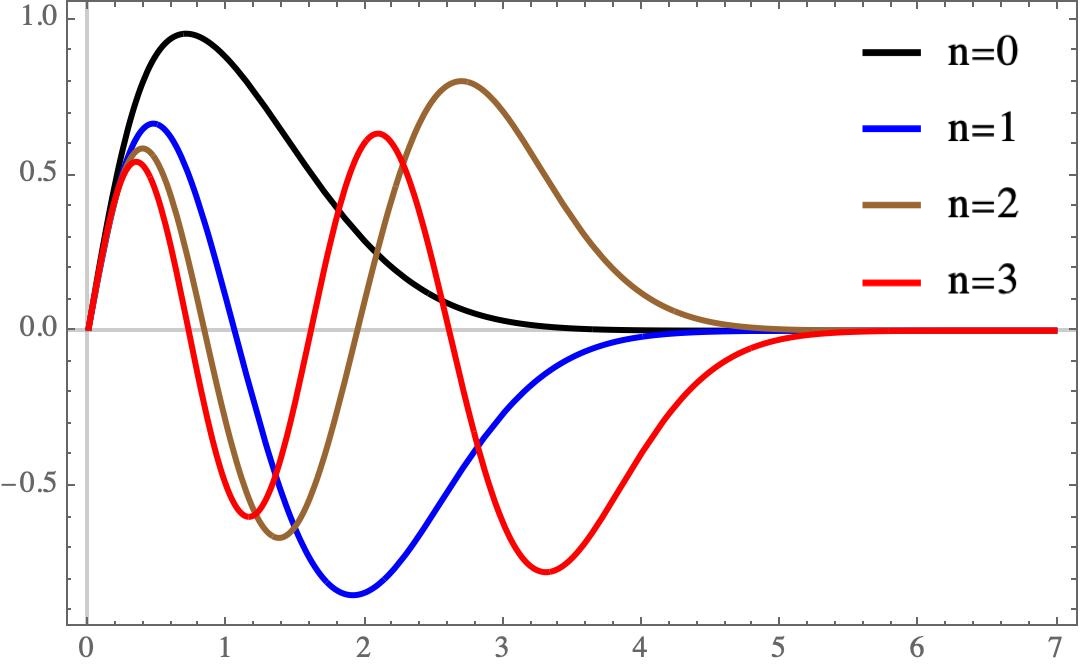}
        \caption{Eigenfunctions $\psi_{h,n}^-(x,m,\alpha)$ vs $x$ for $m=1$ and $\alpha=-\frac{1}{2}$ and $n=0,1,2,3$}
        \label{fig:figure1}
    \end{subfigure}
    \hfill
    \begin{subfigure}[b]{0.48\textwidth}
        \centering
        \includegraphics[width=\textwidth]{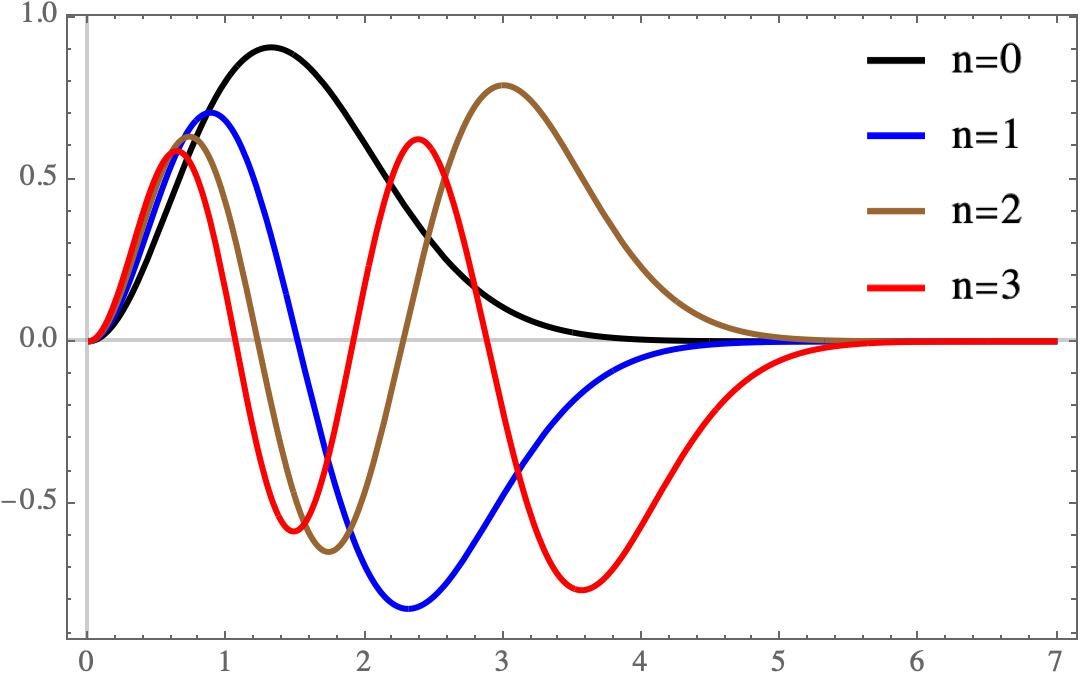}
        \caption{Eigenfunctions $\psi_{h,n}^-(x,m,\alpha)$ vs $x$ for $m=1$ and $\alpha=\frac{1}{2}$ and $n=0,1,2,3$}
        \label{fig:figure2}
    \end{subfigure}
    \caption{Eigenfunctions for RE Half-Line oscillator.}
    \label{fig:main}
\end{figure}

\begin{table}[htp]
    \centering
    \renewcommand{\arraystretch}{1.5} 
    \setlength{\tabcolsep}{1.5em} 
    \begin{tabularx}{\columnwidth}{|c|X|X|}
    \toprule
    {\bf m} &$\mathbf{V_h^-(x,m,-\frac{1}{2})}$&$\mathbf{V_h^-(x,m,\frac{1}{2})}$\\
    \toprule
    0 &$\frac{x^2 \omega_x^2}{4} - \omega_x$ & $\frac{x^2 \omega_x^2}{4} + \frac{2}{x^2} - \omega_x$\\
    \bottomrule
    1 &$\frac{x^2 \omega_x^2}{4} + \frac{4 \omega_x}{x^2 \omega_x + 1} - \frac{8 \omega_x}{\left(x^2 \omega_x + 1\right)^2} - \omega_x$&$\frac{x^2 \omega_x^2}{4} + \frac{4 \omega_x}{x^2 \omega_x + 3} - \frac{24 \omega_x}{\left(x^2 \omega_x + 3\right)^2} + \frac{2}{x^2} - \omega_x$\\
    \bottomrule
    2 &$\frac{x^2 \omega_x^2}{4} + \frac{192 x^2 \omega_x^2}{\left(x^4 \omega_x^2 + 6 x^2 \omega_x + 3\right)^2} + \frac{8 \left(x^2 \omega_x^2 - 3 \omega_x\right)}{x^4 \omega_x^2 + 6 x^2 \omega_x + 3} - \omega_x$ &$\frac{x^2 \omega_x^2}{4} + \frac{320 x^2 \omega_x^2}{\left(x^4 \omega_x^2 + 10 x^2 \omega_x + 15\right)^2} + \frac{8 \left(x^2 \omega_x^2 - 5 \omega_x\right)}{x^4 \omega_x^2 + 10 x^2 \omega_x + 15} + \frac{2}{x^2} - \omega_x$\\
    \bottomrule
    3 &$\frac{x^2 \omega_x^2}{4} + \frac{12 \left(x^4 \omega_x^3 + 45 \omega_x\right)}{x^6 \omega_x^3 + 15 x^4 \omega_x^2 + 45 x^2 \omega_x + 15} - \frac{2160 \left(3 x^4 \omega_x^3 + 10 x^2 \omega_x^2 + 5 \omega_x\right)}{\left(x^6 \omega_x^3 + 15 x^4 \omega_x^2 + 45 x^2 \omega_x + 15\right)^2} - \omega_x$&$\frac{x^2 \omega_x^2}{4} + \frac{12 \left(x^4 \omega_x^3 + 49 \omega_x\right)}{x^6 \omega_x^3 + 21 x^4 \omega_x^2 + 105 x^2 \omega_x + 105} - \frac{1008 \left(11 x^4 \omega_x^3 + 70 x^2 \omega_x^2 + 105 \omega_x\right)}{\left(x^6 \omega_x^3 + 21 x^4 \omega_x^2 + 105 x^2 \omega_x + 105\right)^2} + \frac{2}{x^2} - \omega_x$\\
    \bottomrule
\end{tabularx}
    \caption{Potentials $V^-_h(x,m,\alpha)$ (which are only valid for $x>0$) for different $m$ when $\alpha$ equals $-\frac{1}{2}$ and $\frac{1}{2}$ respectively .}
    \label{Vh-1D-table}
\end{table}

\begin{table}[htp]
    \centering
    \renewcommand{\arraystretch}{1.5} 
    \setlength{\tabcolsep}{1.5em} 
    \begin{tabularx}{\columnwidth}{|c|X|X|}
    \toprule
    {\bf m} &$\mathbf{\psi_{h,n}^-(x,m,-\frac{1}{2})}$&$\mathbf{\psi_{h,n}^-(x,m,\frac{1}{2})}$\\
    \toprule
    0 & $x e^{-\frac{\omega_x  x^2}{4}}L_{n}^{(\frac{1}{2})}\left(\frac{\omega_x}{2} x^2\right) $&$x^2 e^{-\frac{\omega_x  x^2}{4}}L_{n}^{(\frac{3}{2})}\left(\frac{\omega_x}{2} x^2\right)$\\
    \bottomrule
    1 &$x e^{-\frac{\omega_x  x^2}{4}}\left[L_{n-1}^{(\frac{1}{2})}\left( \frac{\omega_x}{2}x^2 \right)+\frac{\left(\omega_x  x^2+3\right) L_n^{(-\frac{1}{2})}\left( \frac{\omega_x}{2}x^2 \right)}{\omega_x  x^2+1} \right]$&$x^2 e^{-\frac{\omega_x  x^2}{4}}\left[L_{n-1}^{(\frac{3}{2})}\left( \frac{\omega_x}{2}x^2 \right)+\frac{\left(\omega_x  x^2+5\right) L_n^{(\frac{1}{2})}\left( \frac{\omega_x}{2}x^2 \right)}{\omega_x  x^2+3}\right] $\\
    \bottomrule
    2 &$x e^{-\frac{\omega_x  x^2}{4}}\Bigg[L_{n-1}^{(\frac{1}{2})}\left( \frac{\omega_x}{2}x^2 \right)+\frac{\left(2\omega_x  x^2 \left(\frac{\omega_x}{2} x^2+5\right)+15\right) L_n^{(-\frac{1}{2})}\left( \frac{\omega_x}{2}x^2 \right)}{2\omega_x  x^2 \left(\frac{\omega_x}{2} x^2+3\right)+3}\Bigg]$ &$x^2 e^{-\frac{\omega_x  x^2}{4}}\Bigg[L_{n-1}^{(\frac{3}{2})}\left( \frac{\omega_x}{2}x^2 \right)+\frac{\left(2\omega_x  x^2 \left(\frac{\omega_x}{2} x^2+7\right)+35\right) L_n^{(\frac{1}{2})}\left( \frac{\omega_x}{2}x^2 \right)}{2\omega_x  x^2 \left(\frac{\omega_x}{2} x^2+5\right)+15}\Bigg]$\\
    \bottomrule
    3 &$x e^{-\frac{\omega_x  x^2}{4}}\Bigg[L_{n-1}^{(\frac{1}{2})}\left( \frac{\omega_x}{2}x^2 \right)+\frac{\left(4\omega_x ^3 x^6+42\omega_x ^2 x^4+105\omega_x  x^2+105\right) L_n^{(-\frac{1}{2})}\left( \frac{\omega_x}{2}x^2 \right)}{4\omega_x ^3 x^6+30\omega_x ^2 x^4+45\omega_x  x^2+15}\Bigg]$ &$x^2 e^{-\frac{\omega_x  x^2}{4}}\Bigg[L_{n-1}^{(\frac{3}{2})}\left( \frac{\omega_x}{2}x^2 \right)+\frac{\left(\omega_x  x^2 \left(2\omega_x ^2 x^4+27\omega_x  x^2+189\right)+315\right) L_n^{(\frac{1}{2})}\left( \frac{\omega_x}{2}x^2 \right)}{4\omega_x ^3 x^6+42\omega_x ^2 x^4+105\omega_x  x^2+105}\Bigg]$\\
    \bottomrule
\end{tabularx}
\caption{Eigenfunctions corresponding to $V_h^-(x,m,\alpha)$ for different $m$ and $\alpha$ equals $-\frac{1}{2}$ and $\frac{1}{2}$ respectively.}
\label{psih-1D-table}
\end{table}


\section{Two-Dimensional Anisotropic Harmonic Oscillator}

In this section, we use the results discussed in the last section and construct
the two-dimensional anisotropic harmonic oscillator RE potentials and the exact eigenfunctions. There are three possible rational extensions of the 2D anisotropic harmonic oscillator, namely
\begin{enumerate}
    \item $2$D-Full-line oscillator
    \item $2$D-Truncated oscillator and
    \item combination of $1$D-Full-line and $1$D-Half-line oscillator.
\end{enumerate}

\subsection{$2$D-Full-line oscillator}
Consider the $2$D-anisotropic harmonic oscillator potential defined on the full
$xy$-plane
\begin{equation}
V^+(x,y)=\frac{1}{4}\omega_x^2 x^2+\frac{1}{4}\omega_y^2 y^2; \quad -\infty<x<\infty,-\infty<y<\infty
\end{equation}
with the eigenfunctions and the eigenvalues
\begin{eqnarray}\label{FLO2D:psi+}
\psi_{n_1,n_2}^+(x,y)&\propto& \psi_{n_1}^+(x)\psi_{n_2}^+(y)\nonumber\\
&\propto&e^{-\frac{\omega_x x^2+\omega_y y^2}{4} } H_{n_1}\left(\sqrt{\frac{\omega_x}{2}} x\right) H_{n_2}\left(\sqrt{\frac{\omega_y}{2}} y\right)\; 
\quad n_1,n_2=0,1,2,\cdots, 		
\end{eqnarray}
and
\begin{eqnarray}
E^+_{n_1,n_2}=E^+_{n_1}+E^+_{n_2}
\end{eqnarray}
respectively where $E^{+}_{n}$ is given by Eq. (\ref{HLO:E-}). The seedless function along each axis can be constructed by replacing the frequencies in the eigenfunctions of conventional potential (\ref{FLO2D:psi+}) with imaginary frequencies and the integers $n_1$ and $n_2$ are replaced by $m_1$ and $m_2$ respectively. This way one gets $(m_1,\;m_2)$-dependent seedless function defined as
\begin{align}
\phi_{m_1,m_2}(x,y)=\prod_{k=1}^{2}\phi_{m_k}(x_k)\quad\text{where}\quad (x_1,x_2)\rightarrow (x,y)\,.
\end{align}
In this way, we get the seedless function 
\begin{equation}\label{FLO2D:phi+}
	\phi_{m_1,m_2}(x,y) \propto e^{\frac{\omega_x\;x^2+\omega_y\; y^2}{4}} H_{m_1}\left(i\sqrt{\frac{\omega_x}{2}} x\right)H_{m_2}\left(i\sqrt{\frac{\omega_y}{2} } y\right)
\end{equation}
and this gives \(Q_2=0\) (see the Eq. (\ref{F:Q}) in the Appendix B) and 
therefore the SUSY partners of \(V^+(x)\), i.e. the RE potentials 
$V^-(x,y,m_1,m_2)$ are given by adding the individual partner potentials as 
given in (\ref{FLO:V-}). In particular
\begin{equation}\label{2D:V-}
V^-(x,y,m_1,m_2)=V^-(x,m_1)+V^-(y,m_2)\,
\end{equation}
where $m_1$ and $m_2$ are both positive even integers. The ground and the 
excited state eigenfunctions are given by 
\begin{align} \label{2D:psi-}
    \begin{split}
        \psi^-_{0,0}(x,y,m_1,m_2)&=\phi_{m_1}^{-1}(x)\phi_{m_2}^{-1}(y), \\
        \mbox{and} \quad  \psi^-_{n_1+1,n_2+1}(x,y,m_1,m_2)&=\psi^-_{n_1+1}(x,m_1)\psi^-_{n_2+1}(y,m_2)\;,\qquad n_1,n_2=0,1,2,\cdots\\
    \end{split}
\end{align}
respectively while the energy eigenvalues are given by
\begin{align}
 E_{n_1+1,n_2+1,m_1,m_2}^-&=\left[\left(n_1+m_1+1\right)\omega_x
+\left(n_2+m_2+1\right)\omega_y \right] \quad \text{and}
\quad E_{0,0,m_1,m_2}^-=0.
\end{align}
Notice that the spectrum of the extended potential is strictly isospectral to the conventional starting potential and hence the form of the degeneracy will be identical to that of the $2D$-anisotropic harmonic oscillator potential (see for example \cite{louck1973canonical}). Notice that the degeneracy occurs if the ratio of the two frequencies $\omega_x$ and $\omega_y$ is a rational number.

\subsection{2D-Half-Line Oscillator}
The half-line oscillator is defined in the region \((x,y) \in (0,\infty) \times
(0,\infty)\). The potential is given by
\[ V^+_h(x,y) = 
 \begin{cases} 
 \frac{1}{4}\omega_x^2 x^2 + \frac{1}{4}\omega_y^2 y^2 & \text{for} (x,y) \in (0,\infty) \times (0,\infty), \\
 \infty & \text{otherwise}
 \end{cases}
\]
 Using the results obtained for the one-dimensional case in the previous 
 section i.e. equation (\ref{HLO:psi+}), the eigenfunctions for this potential 
 are given by
\begin{equation}\label{2DHLO:psi+}
    \psi_{h,n_1,n_2}^+(x,y,\alpha,\beta)\propto  x^{\alpha +\frac{1}{2}}y^{\beta +\frac{1}{2}} e^{-\frac{\omega_xx^2+\omega_y y^2}{4}} L_{n_1}^{(\alpha) }\left(\frac{\omega_x}{2} x^2\right)L_{n_2}^{(\beta)}\left(\frac{\omega_y}{2} y^2\right); \quad n_1, n_2=0,1,2,\cdots
\end{equation} 
where, $\beta=\pm\frac{1}{2}$ is another parameter for the potential defined 
along the $y$ direction. The seedless solution 	
$\phi_{h,m_1,m_2}(x,y,\alpha,\beta)$ corresponding to this potential can be 
easily constructed by replacing $(n_1,n_2)\rightarrow (m_1,m_2)$ and 
$(\omega_x,\omega_y)\rightarrow (-\omega_x, -\omega_y)$ in 
$\psi_{n_1,n_2}^+(x,y,\alpha,\beta)$ and is given by
\begin{equation}\label{2DHLO:phi}
    \phi_{h, m_1,m_2}(x,y,\alpha,\beta)\propto x^{\alpha +\frac{1}{2}}y^{\beta +\frac{1}{2}} e^{\big(\frac{\omega_x x^2+\omega_y y^2}{4}\big)} 
L_{m_1}^{(\alpha)}\left(-\frac{\omega_x}{2} x^2\right)L_{m_2}^{(\beta)}\left(-\frac{\omega_y}{2}  y^2\right); \qquad m_1,m_2=0,1,2,\cdots \,.
\end{equation}
The corresponding energy eigenvalues of the Hamiltonian $H_h^+(x,y)$ is obtained using
$V^+_h(x,y)-\epsilon_{h,m_1,m_2}$ as
\begin{equation}\label{2DHLO:E+}
    E^+_{h,n_1,n_2,\alpha,\beta}=[(2n_1+\alpha+1)\omega_x+(2n_2+\beta+1)\omega_y]-\epsilon_{h,m_1,m_2},
\end{equation}
where $\epsilon_{h,m_1,m_2}=-[(2m_1+\alpha+1)\omega_x+(2m_2+\beta+1)\omega_y]$ 
is the factorization energy corresponding to the seedless function $\phi_{h,m_1,m_2}(x,y,\alpha,\beta)$. There are four possible combinations of the parameters
$\alpha$ and $\beta$ (i.e. $\alpha=\beta=\pm 1/2$ and $\alpha=1/2, \beta=-1/2$ 
and $\alpha=-1/2, \beta= 1/2$), out of which the case $\alpha=1/2, \beta=-1/2$ 
and $\alpha=-1/2, \beta=1/2$ are equivalent. Thus effectively we have three 
forms of the RE potentials which we now discuss one by one. 

{\bf Case (a)} For $\alpha=\beta=1/2$
    
The form of the extended potential, the corresponding eigenfunctions and the 
energy eigenvalues are given by 
\begin{align}
V_h^-(x,y,m_1,m_2,\frac{1}{2}) &= V_h^-(x,m_1,\frac{1}{2}) + V_h^-(y,m_2,\frac{1}{2}) \\
\psi_{h,n_1,n_2}^-(x,y,m_1,m_2,\frac{1}{2})&\propto\psi^-_{n_1}(x,m_1,\frac{1}{2})\psi^-_{n_2}(y,m_2,\frac{1}{2})\\
\mbox{and} \quad	
E_{h,n_1,n_2}^-(m_1,m_2,\frac{1}{2})&=\left[\left(2n_1+2m_1+3\right)\omega_x 
+\left(2n_2+2m_2+3\right) \omega_y \right]\,
\end{align}
respectively. Here the potentials $V_h^-(x,m_1,\frac{1}{2})$ and  
$V_h^-(y,m_2,\frac{1}{2})$ and the eigenfunctions 
$\psi^-_{n_1}(x,m_1,\frac{1}{2})$ and $\psi^-_{n_2}(y,m_2,\frac{1}{2})$ can be 
easily obtained from Eqs. (\ref{HLO:V-alpha+}) and (\ref {HLO:psi-}) 
respectively. In the special case of the isotropic oscillator (i.e. $\omega_x =
\omega_y$) the results for the RE potential (where $m_1=m_2=m$(say)), are the 
same as those obtained in \cite{yadava2017rationally} for $D=2$ and are valid 
for all positive values of $m$. 	

{\bf Case (b)}  For \(\alpha = \beta=-1/2\)

In this case, the form of potential with the corresponding energy 
eigenfunctions and the energy eigenvalues are 
\begin{align}
    V_h^-(x,y,m_1,m_2,-\frac{1}{2}) &= V_h^-(x,m_1,-\frac{1}{2}) + V_h^-(y,m_2,-\frac{1}{2}) \\
    \psi_{h,n_1,n_2}^-(x,y,m_1,m_2,-\frac{1}{2})&\propto\psi^-_{h,n_1}(x,m_1,-\frac{1}{2})\psi^-_{h,n_2}(y,m_2,-\frac{1}{2})\\
    \mbox{and} \quad	
E_{h,n_1,n_2}^-(m_1,m_2,\frac{1}{2})&=\left[\left(2n_1+2m_1+1\right)\omega_x
+\left(2n_2+2m_2+1\right) \omega_y \right]\,.
\end{align}
Here the potentials $V_h^-(x,m_1,-\frac{1}{2})$ and  
$V_h^-(y,m_2,-\frac{1}{2})$ and the eigenfunctions 
$\psi^-_{n_1}(x,m_1,-\frac{1}{2})$ and $\psi^-_{n_2}(y,m_2,-\frac{1}{2})$ can 
be easily obtained from Eqs. (\ref{HLO:V-alpha-}) and (\ref {HLO:psi-}) 
respectively. 

{\bf Case (c) }  For $\alpha = \frac{1}{2}$ and $\beta=-\frac{1}{2}$

Similar to the above cases, the potential, the eigenfunctions and the energy 
eigenvalues are given by 
\begin{align}
    V_h^-(x,y,m_1,m_2,\frac{1}{2},-\frac{1}{2}) &= V_h^-(x,m_1,\frac{1}{2}) + V_h^-(y,m_2,-\frac{1}{2}) \\
    \psi_{h,n_1,n_2}^-(x,y,m_1,m_2,\frac{1}{2},-\frac{1}{2})&\propto\psi^-_{h,n_1}(x,m_1,\frac{1}{2})\psi^-_{h,n_2}(y,m_2,-\frac{1}{2})\\
    \mbox{and} \quad	
E_{h,n_1,n_2}^-(m_1,m_2,\frac{1}{2},-\frac{1}{2})&=\left[\left(2n_1+2m_1+3\right)\omega_x +\left(2n_2+2m_2+1\right)\omega_y\right]\,
\end{align}
The form of the degeneracy in this case is similar to the full line case and
hence identical to that of the $2D$-anisotropic harmonic oscillator potential 
(see for example \cite{louck1973canonical}). Notice that the degeneracy occurs 
when the ratio $\frac{\omega_x}{\omega_y}$ is a rational number.

\subsection{One Full-Line and One Half-Line Oscillator}

This two dimensional anisotropic harmonic oscillator is defined with one 
coordinate spanning the entire real line 
(say along the $x-$axis) and the other truncated at zero (say along the 
$y-$axis) as
\[ V^+_{fh}(x,y) = 
\begin{cases} 
\frac{1}{4}\omega_x^2 x^2 + \frac{1}{4}\omega_y^2 y^2 & \text{for } y > 0; -\infty<x<\infty, \\
\infty & \text{for } y \le 0
\end{cases}
\]
The rational extension for this combined full-line and a half-line oscillators  can be easily constructed by using the known results as given by Eqs. 
(\ref{FLO:V-}), (\ref{HLO:V-alpha+}) and (\ref{HLO:V-alpha-}) respectively. 
As already discussed in the previous section, while in the case of the full 
line oscillator there is one extra bound state, however in the case of the 
half-line oscillator (which is $\beta$ dependent), there is no extra bound 
state. The general form of the extended potential is given as 

\begin{equation}\label{GFHP-}
	V^-_{fh}(x,y,m_1,m_2,\beta)=V^-(x,m_1)+V^-_h (y,m_2,\beta),
\end{equation}
here $m_1=0,2,4,\cdots$, $m_2=0,1,2,3,\cdots$ and $\beta=\pm 1/2$. The 
corresponding ground and the excited state eigenfunctions are  
\begin{equation}
\psi^-_{fh,0,0}(x,y,m_1,m_2,\beta)\propto\psi^-_0 (x,m_1)\psi^-_{h,0}(y,m_2,\beta)
\end{equation}
and
\begin{equation}
\psi^-_{fh,n_1+1,n_2}(x,y,m_1,m_2,\beta)\propto\psi^-_{n_1+1} (x,m_1)\psi^-_{h,n_2}(y,m_2,\beta);	
\end{equation}
respectively. Here $n_1=0,1,2,\cdots$ and $n_2=1,2,3,\cdots$. The corresponding 
expressions for the ground and the excited state energy eigenvalues are
\begin{equation}\label{fhg}
	E^-_{fh,0,0}(m_2,\beta)=2(m_2+\beta+1)\omega_y.
\end{equation}
and
\begin{eqnarray}\label{fhexct}
	E^-_{fh,n_1+1,n_2}(m_1,m_2,\beta)&=& E^+_{n_1}(m_1)+E^+_{h,n_2}(m_2,\beta)\nonumber\\
	&=& (n_1+m_1+1)\omega_x+2(n_2+m_2+\beta+1)\omega_y		
\end{eqnarray}
respectively. The explicit expressions for the extended potentials and the corresponding energy eigenvalues and eigenfunctions in case $\beta=\pm1/2$ are as follows:

{\bf Case (a)} For \(\beta = \frac{1}{2}\)
 
In this case, we use the expressions of $V^-(x,m_1)$, $V_h^-(y,m_2,\frac{1}{2})$ and $V^-_h (y,m_2,-1/2)$ from Eqs. (\ref{FLO:V-}), (\ref{HLO:V-alpha+}) and (\ref{HLO:V-alpha-}) respectively, and get
\begin{align}
    V^-_{fh}(x,y,m_1,m_2,1/2) &= V^-(x,m_1)+V_h^-(y,m_2,\frac{1}{2})
\end{align}
The corresponding ground and the excited state eigenfunctions and the the 
corresponding energy eigenvalues are given by

\begin{align}
    \psi^-_{fh,0,0}(m_1,m_2,\frac{1}{2}) &\propto \psi^-_0 (x,m_1)\psi^-_{h,0}(y,m_2,\frac{1}{2})
    \\ 
    \psi^-_{n_1+1,n_2}(m_1,m_2,\frac{1}{2}) &\propto \psi^-_{n_1+1} (x,m_1)\psi^-_{h,n_2}(y,m_2,\frac{1}{2})
\end{align}
and 
\begin{align} 
    E^-_{0,0}(m_1,m_2,\frac{1}{2}) &=\left(2m_2+3\right)\omega_y\\
    E^-_{n_1+1,n_2}(m_1,m_2,\frac{1}{2}) &= \left(n_1+2m_1+1\right)\omega_x+ \left(2n_2+2m_2+3\right)\omega_y
\end{align}
respectively.

{\bf Case (b)} For \(\beta = -\frac{1}{2}\)

Here the form of the extended potential is given by
\begin{align}
	V^-_{fh}(x,y,m_1,m_2,-1/2) &= V^-(x,m_1)+V_h^-(y,m_2,-\frac{1}{2})
\end{align}
The corresponding ground and the excited state eigenfunctions and the energy 
eigenvalues are

\begin{align}
	\psi^-_{fh,0,0}(m_1,m_2,-\frac{1}{2}) &\propto \psi^-_0 (x,m_1)\psi^-_{h,0}(y,m_2,-\frac{1}{2})  \\ 
	\psi^-_{n_1+1,n_2}(m_1,m_2,-\frac{1}{2}) &\propto \psi^-_{n_1+1} (x,m_1)\psi^-_{h,n_2}(y,m_2,-\frac{1}{2})
\end{align}
and 
\begin{align} 
	E^-_{0,0}(m_1,m_2,-\frac{1}{2}) &=\left(2m_2+1\right)\omega_y\\
	E^-_{n_1+1,n_2}(m_1,m_2,-\frac{1}{2}) &= \left(n_1+2m_1+1\right)\omega_x+ \left(2n_2+2m_2+1\right)\omega_y
\end{align}
respectively. 

One point worth mentioning here. As seen above, when both the oscillators are on the full line or both on the half line, the degeneracy for given $m_1,m_2$ essentially comes from the factor $n_1\omega_x+n_2\omega_y$. However, when one oscillator is on the full line and the other is on the half-line then for a given $m_1,m_2$, the degeneracy essentially comes from the factor $n_1\omega_x+2n_2\omega_y$, which is different from the two above cases.  As a result, for a given rational value of $\frac{\omega_x}{\omega_y}$, the degeneracy will be different in this case in contrast to the other two cases.


\section{Three-Dimensional Anharmonic Harmonic Oscillator}

Similar to the 2D anisotropic case, generalization to arbitrary $D$ dimensions 
is straight forward. For example, using the seedless function for each 
potential \(V^+(x_k)\) along a given axis (full line or half line), one can 
construct a $(m_1,m_2,m_3)$-dependent three dimensional RE anisotropic harmonic
oscillator potential \(V^-(x,y,z,m_1,m_2,m_3)\) by starting from the known 
anisotropic three dimensional potential \(V^+(x,y,z)\), whose expression is 
given by (\ref{F:V-}) with zero value of \(Q_3\) in (\ref{F:Q}). The 
corresponding eigenfunctions and the energy eigenvalues are given by 
(\ref{F:psi-}) and (\ref{F:E-}) respectively. 

In the case of all three 
half-line oscillators, with all three frequencies equal, there are two possible
forms of the extended potentials corresponding to \(\alpha=\pm\frac{1}{2}\). 
For $\alpha=\frac{1}{2}$ the extended potential obtained in polar coordinates 
reduces to the same form as already obtained in \cite{yadava2017rationally} for
$D$-dimensional isotropic harmonic oscillator with $D=3$ and $l=1$, except for 
an additional constant of $-\frac{3}{2}\omega$ in \cite{yadava2017rationally}. 
But for $\alpha=-1/2$, we get another form of the extended isotropic potential.
The form of this potential with their solutions can easily be obtained using 
the results obtained for the one-dimensional case and using Eqs. 
(\ref{HLO:V-alpha-}), (\ref{HLO:psi-}) and (\ref{HLO:E-}). It is then 
worthwhile to consider the problem of the 3-dimensional anisotropic oscillator
in case two of the three frequencies are the same and obtain their rational 
extension.

\subsection{Full-Line Oscillator with two equal frequencies}

Consider the following anisotropic potential defined on the full-line
\begin{equation}
V^+(x,y,z)=\frac{1}{4}\omega^2 (x^2+y^2)+\frac{1}{4}\omega_z^2 z^2; \quad -\infty<x,y,z<\infty
\end{equation}
where $\omega_x=\omega_y=\omega$. In this case the above potential can be written in the $r-z$ co-ordinates as
\begin{equation}
  V^+(r,z)=\frac{1}{4}\omega^2 r^2+\frac{1}{4}\omega_z^2 z^2\,
\end{equation}
where $r^2=x^2+y^2$. 
The form of the Schr\"odinger equation in the cylindrical co-ordinates ($r,\phi,z$) is
\begin{equation}\label{cylinderical-SE}
  \left[-\left(\frac{1}{r}\frac{\partial }{\partial r}r\frac{\partial }{\partial r}+\frac{1}{r^2}\frac{\partial^2 }{\partial \phi^2}+\frac{\partial^2 }{\partial z^2}\right)+V^+(r,z)\;\right]\psi^+(r,\phi,z)=E\psi^+ (r,\phi,z),
\end{equation}
where $\psi^+ (r,\phi,z)$ is the eigenfunction and $E$ is the energy eigenvalue corresponding to the potential $V^+(r,z)$. One can easily solve the 
Eq. (\ref{cylinderical-SE}) by assuming 
\begin{equation}\label{rphiz}
\psi^+ (r,\phi,z)=\frac{e^{i\gamma \phi} \zeta(r,z)}{r^{\frac{1}{2}}}\,.
\end{equation}
where
\begin{equation}
\zeta(r,z) = R(r) Z(z)\,.
\end{equation}
One can show that in this case $\zeta(r,z)$ satisfies the equation
\begin{equation}
\left[-\frac{\partial^2}{\partial r^2}-\frac{\partial^2}{\partial z^2}
+V^+_{eff}(r,z)\right]\zeta(r,z)=E\;\zeta(r,z),
\end{equation}
where the effective potential $V^+_{eff}(r,z)$ is given by
\begin{equation}
V^+_{eff}(r,z)=V^+(r,z)+\frac{\gamma^2-1/4}{r^2}\quad\text{with}\quad  
\gamma=0,\pm1,\pm2,\cdots \,.
\end{equation} 
The functions $R(r)$ and $Z(z)$ can be shown to be
\begin{equation}
R(r)\rightarrow R_{n_1}(r)\propto r^{|\gamma|+\frac{1}{2}}e^{-\frac{\omega}{4}r^2}L_{n_1}^{|\gamma|}(\frac{1}{2}\omega r^2); n_1=0,1,2,\cdots
\end{equation}
and 
\begin{equation}
Z(z)\rightarrow Z_{n_2}(z)\propto e^{-\frac{\omega_z z^2}{4}}H_{n_2}(\sqrt{\frac{\omega_z}{2}} z); n_2=0,1,2,\cdots
\end{equation}
respectively. The corresponding energy eigenvalues $E$ turn out to be
\begin{equation}
  E\rightarrow E_{n_1,n_2}=(2n_1+|\gamma|+1)\omega+\left(n_2+\frac{1}{2}\right)
\omega_z\,.
\end{equation}
Thus the energy eigenvalues for the Hamiltonian 
$H^+(r,z)=V^+(r,z)-\epsilon_{m_1,m_2}$ are
\begin{equation}\label{CYL:E+}
 E^+_{n_1,n_2}=E_{n_1,n_2}-\epsilon_{m_1,m_2}=2(n_1+m_1+|\gamma|+1)\omega
+(n_2+m_2+1)\omega_z.
\end{equation}
Now one can easily construct the corresponding RE potential $V_{eff}^+(r,z)$ 
by defining the seedless functions $\phi_{m_1}(r)$ and $\phi_{m_2}(z)$, by 
replacing $n_1\rightarrow m_1$, $\omega\rightarrow-\omega$  and 
$n_2\rightarrow m_2$, $\omega_z\rightarrow-\omega_z$ in the expressions for 
$R_{n_1}(r)$ and $Z_{n_2}(z)$ respectively. We obtain 
\begin{align}
\begin{split}
    \phi_{m_1}(r)&\propto r^{|\gamma| +\frac{1}{2}} e^{\frac{\omega}{4}r^2} L_{m_1}^{|\gamma| }\left(-\frac{1}{2}\omega  r^2\right);\qquad m_1=0,1,2\cdots\\
    \phi_{m_2}(z)&\propto e^{\frac{\omega_z}{4} z^2} \mathcal{H}_{m_2}\left(\sqrt{\frac{\omega_z}{2} } z\right);\qquad m_2=0,2,4\cdots\quad\text{Since $z$ axis is full line.}
\end{split}
\end{align}
Thus the combined seedless function corresponding to the wavefunction 
$\zeta(r,z)$ will be 
\begin{equation}
\phi_{m_1,m_2}(r,z)=\phi_{m_1}(r)\phi_{m_2}(z)
\end{equation} 
and the operators $A$ and $A^{\dagger}$ in term of $\phi_{m_1,m_2}(r,z)$ are given by
\begin{align}
	\begin{split}
		A&=\frac{\partial}{\partial r}+\frac{\partial \log(\phi_{m_1,m_2}(r,z))}{\partial r}+i\frac{\partial}{\partial z}+i \frac{\partial\log\phi_{m_1,m_2}(r,z)}{\partial z}\\
	\mbox{and} \quad	A^{\dagger}&=-\frac{\partial}{\partial r}+\frac{\partial \log(\phi_{m_1,m_2}(r,z))}{\partial r}+i\frac{\partial}{\partial z}-i \frac{\partial\log\phi_{m_1,m_2}(r,z)}{\partial z}
	\end{split}
\end{align}
respectively. The RE potential $V_{eff}^-(m_1,m_2,r,z)$ with the corresponding
eigenfunctions $\zeta_{n_1,n_2}^-(r,z,m_1,m_2)$ and energy eigenvalues 
$E^-_{n_1,n_2}(m_1,m_2)$ corresponding to the potential $V_{eff}(r,z)$ are
\begin{align}
    V_{eff}^-(m_1, m_2, r, z) &= \frac{1}{4}\omega^2 r^2 + \frac{\gamma^2 - \frac{1}{4}}{r^2} + \frac{2 | \gamma | }{r^2}-(2 m+1) \omega +\frac{1}{r^2} \nonumber\\
    &\quad + \frac{2 r^2 \omega ^2 L_{m_1-1}^{(| \gamma | +1)}\left(-\frac{\omega r^2}{2}\right){}^2}{L_{m_1}^{(| \gamma |) }\left(-\frac{\omega r^2}{2}\right){}^2} \nonumber\\
    &\quad + \frac{\omega  \left(\left(2 | \gamma | +r^2 \omega \right) L_{m_1-1}^{(| \gamma | +1)}\left(-\frac{\omega r^2}{2}\right)-r^2 \omega  L_{m-2}^{(| \gamma | +2)}\left(-\frac{\omega r^2}{2}\right)\right)}{L_{m_1}^{(| \gamma | )}\left(-\frac{\omega r^2}{2}\right)}\nonumber\\
    &\quad + \frac{1}{2}\omega_z^2 z^2 - 2\left[\frac{\mathcal{H}_{m_2}''\left(\sqrt{\frac{\omega_z}{2}} z\right)}{\mathcal{H}_{m_2}\left(\sqrt{\frac{\omega_z}{2}} z\right)}-\left[\frac{\mathcal{H}_{m_2}'\left(\sqrt{\frac{\omega_z}{2}} z\right)}{\mathcal{H}_{m_2}\left(\sqrt{\frac{\omega_z}{2}} z\right)}\right]^2+\frac{\omega_z}{2}\right] \nonumber\\
    \zeta_{n_1, n_2+1}^-(r, z, m_1, m_2) &\propto r^{|\gamma| + \frac{3}{2}} e^{-\frac{\omega r^2}{4}} \frac{\hat{L}_{n_1, m_1}^{\gamma}\left(\frac{1}{2}\omega r^2 \right)}{L_{m_1}^{\gamma}\left(-\frac{1}{2}\omega r^2 \right)}\; e^{-\frac{\frac{\omega_z}{2} z^2}{2}} \frac{\hat{H}_{n_z+1, m_2}\left(\sqrt{\frac{\omega_z}{2}} z \right)}{\mathcal{H}_{m_2}\left(\sqrt{\frac{\omega_z}{2}} z \right)} \nonumber\\
    E^-_{n_1, n_2+1}(m_1, m_2) &= 2(n_1 + m_1 + |\gamma| + 1)\omega + \left(n_2 + m_2 + 1\right)\omega_z
\end{align}
with the ground state eigenfunctions and the energy eigenvalues 
\begin{eqnarray}
\zeta_{0,0}^-(r,z,m_1,m_2)&\propto& r^{|\gamma| +\frac{3}{2}}  \frac{e^{-\frac{\omega r^2}{4}}L_{\text{m1}}^{| \gamma | +1}\left(\frac{1}{2}\omega \left(-r^2\right)\right)}{L_{m_1}^{|\gamma |}\left(-\frac{1}{2}\omega r^2 \right)}\times
\frac{e^{-\frac{\omega_z\;  z^2}{4}}}{\mathcal{H}_{m_2}^{}\left(\sqrt{\frac{\omega_z}{2}}  z \right)}  \nonumber\\
\mbox{and} \quad E^-_{0,0,m_1,m_2}&=&2(m_1+|\gamma|+1)\omega
\end{eqnarray}
respectively, where, the prime over pseudo-Hermite polynomial 
$\mathcal{H}_m(\sqrt{\frac{\omega_z}{2}} z)$ denotes derivatives with respect to $z$. The RE potential is isospectral but not strictly isospectral to $V^{+}$ as there is additional bound state along $z$-axis. The form of the degeneracy is therefore the same for the RE potential and $V^{+}$. The degeneracy in the RE case (for given $m_1, m_2$) occurs when the ratio of $\omega$ and $\omega_z$ is a rational number.
The form of the degeneracy in this case is similar to the full line case and
hence identical to that of the $2D$-anisotropic harmonic oscillator potential 
(see for example \cite{louck1973canonical}). Notice that the degeneracy occurs 
when the ratio $\frac{\omega_x}{\omega_y}$ is a rational number.

\section{Conclusions and Open Problem}
In this paper, we began with the one-dimensional harmonic oscillator and 
derived the rational extension for the potential on the half-line. We also 
obtained the corresponding eigenfunctions, which are expressed in terms of 
Laguerre polynomials. The solutions are dependent on the parameter 
\(\alpha=\pm\frac{1}{2}\). While the result for \(\alpha = +\frac{1}{2}\) is 
well-known, however the case for \(\alpha = -\frac{1}{2}\) is new and novel.
We then extended these results to the two-dimensional anisotropic harmonic 
oscillator, discussing the RE potentials in the three possible cases, i.e. the 
full-line oscillator along both the axes, the half-line oscillator along both 
the axes, and a combination of
the full-line oscillator on one axes and the half-line oscillator on the other.
For the half-line oscillator, we found that no additional bound states exist, 
whereas for the full-line oscillator, an additional bound state with zero 
energy appears. The rational extension of the anharmonic harmonic oscillator to
higher dimensions is easily done using SUSY in two, three and higher 
dimensions. Finally, we also considered one interesting case of 
three-dimensional anisotropic oscillator, where we constructed the RE potential
in the case of the full line oscillator system with two equal frequencies and 
obtained the solution in terms of exceptional Laguerre and Hermite polynomials 
using cylindrical coordinates. 

Now that one has obtained the RE potentials for the anisotropic harmonic 
oscillator potential in two and higher dimensions, one obvious question is
can we add some perturbation to these potentials and still obtain the 
corresponding rational extension or can we generate an extended family of potentials with some perturbation corresponding to these potentials? We hope to address this question in the 
near future. 

\section*{Acknowledgement}
RKY acknowledges Sido Kanhu Murmu University, Dumka, for the grant sanctioned through university letter No. SKMU/CCDC/349, provided under the research project of the State Higher Education Council, Ranchi, Jharkhand. AK is gratefel to Indian National Science Academy (INSA) for awarding INSA Honorary Scientist position at Savitribai Phule Pune University.

\appendix

\renewcommand{\theequation}{A.\arabic{equation}} 
\setcounter{equation}{0}

\section*{Appendix A: SUSY in  One Dimension}

The Schr\"{o}dinger equation (in units $\hbar=2M=1$, where \( M \) is the mass) for a system with potential \( V(x) \) is represented by the equation 
\begin{equation} \label{F:SE}
    \left[-\nabla^2 + V(x)\right]\psi_n(x) = E\psi_n(x)
\end{equation}
or, compactly, 
\begin{equation}\label{F:SE-compact}
    H\psi_n(x) = E\psi_n(x),
\end{equation}
where \(H\) is the second-order Hamiltonian operator. One can factorize the 
Hamiltonian into two linear operators \( A \) (annihilation operator) and 
\( A^\dagger \) (creation operator) to get two possible Hamiltonians 
\( H^{\pm} \) as the operators are non-commutative in general. 
\begin{align}\label{F:A-1D}
    \begin{split}
        A &= \frac{d}{dx} + W(x)\\
        \text{and}\quad A^\dagger &= -\frac{d}{dx} + W(x)  
    \end{split}
\end{align}
where \( W(x) \) is the superpotential. Depending on the order of the operators, the two Hamiltonians are given by
\begin{equation}\label{F:H-1D}
    H - \epsilon=
    \begin{cases}
        A^\dagger A=-\frac{d^2}{dx^2} + V^-(x) - \epsilon, & \text{say } H^-\\
        A A^\dagger=-\frac{d^2}{dx^2} + V^+(x) - \epsilon, & \text{say } H^+
    \end{cases}
\end{equation}
where \( V^{\mp}(x) \) are the partner potentials having the expressions
\begin{equation}\label{F:V-1D}
    V^\mp(x)=W(x)^2\mp W(x)'+\epsilon
\end{equation}
and $\epsilon$ is the factorization energy assumed less than or equal to the ground state energy $E_0$ of $V^+(x)$.
\begin{equation}
    \left[-\frac{d^2}{dx^2}+V^+(x)\right]\psi_0^+(x)=E_0^+\psi_0^+(x)
\end{equation}
and seedless eigenfunction satisfy the following equation
\begin{equation}
    \left[-\frac{d^2}{dx^2}+V^+(x)\right]\phi_m(x)=\epsilon_m\phi_m(x)
\end{equation}
Let us call the eigenfunctions corresponding to Hamiltonian \( H^-(x,m) \) which is $m$-dependent as \( \psi^-(x,m) \) and to Hamiltonian \( H^+(x,m) \) as \( \psi^+(x) \) as it is $m$-independent. The 
Hamiltonian $H^-$ is a factorized Hamiltonian giving zero on operation to the 
ground state eigenfunction $\psi_0^-(x,m)$ of $V^-(x,m)$.
We will focus on the Hamiltonian $H^-$ in this paper assuming $V^+$ (which is $m$-independent) potential 
is known in advance and $V^-$ is constructed using the superpotential \( W(x) \) which is given in terms of the seedless eigenfunction \( \phi_m(x) \) as
\begin{align}
    W(x)&=\begin{cases}\label{F:W-1D}
        -\frac{d}{dx}\ln[\phi_m(x)], & \text{for } \epsilon_m=E_0\\
        +\frac{d}{dx}\ln[\phi_m(x)], & \text{for } \epsilon_m<E_0
    \end{cases}
\end{align}
The sign in expression (\ref{F:W-1D}) varies due to changes in the ground state wavefunction's dependence, being proportional to \( \phi_m(x) \) if \( \phi_m(x) \) is normalizable, or to \( \phi_m^{-1}(x) \) if \( \phi_m^{-1}(x) \) is normalizable. Following cases \cite{marquette2013two} arise:
\begin{itemize}
    \item When $\epsilon_m=E_o^+$, then \( \phi_m(x) \) is the ground state eigenfunction of \( V^-(x,m) \), and the partner potential \( V^+(x) \) is isospectral to the former with only the ground state energy removed. The expression of extra bound state eigenfunction, which is also the ground state eigenfunction of \( V^-(x,m) \), is given by
    \begin{equation}\label{F:GS-I}
        \psi^-_0(x,m)\propto\phi_m(x)
    \end{equation}

    \item When $\epsilon_m<E_o^+$, two possibilities arise:
    \begin{itemize}
        \item If \( \phi_m^{-1}(x) \) is normalizable then the partner potential \( V^-(x,m) \) has an extra bound state which is the ground state with zero energy given by
        \begin{equation}\label{F:GS-III}
            \psi^-_0(x,m) \propto\phi^{-1}_m(x)
        \end{equation} 

        \item If neither \( \phi_m(x) \) nor \( \phi^{-1}_m(x) \) is normalizable, then the partner potential \( V^-(x,m) \) is strictly isospectral to \( V^+(x) \) and therefore no extra bound state is present.
    \end{itemize}
\end{itemize}
The eigenfunctions of the partner Hamiltonian $H^-$ can be obtained using interwing relation \cite{fellows2009factorization,marquette2013two} of linear operators defined in (\ref{F:A-1D}) between the partner Hamiltonians in (\ref{F:H-1D}) as
\begin{align}
    \left(A\;A^\dagger\right) A\;\psi^-&=H^+A\psi^-=E^+\frac{1}{C_-}\psi^+\label{F:int1}\\
    \left(A^\dagger A\right)\;A^\dagger\;\psi^+&=H^-A^\dagger\psi^+=E^-\frac{1}{C_+}\psi^-\label{F:int2} 
\end{align}
where \( C_- \) and \( C_+ \) are normalization constants equal to \( \frac{1}{\sqrt{E^+_{n,m}}} \) and \( \frac{1}{\sqrt{E^-_{n+1,m}}} \) respectively which are related to the eigenvalues of Hamiltonian $H^+$ and $H^-$ respectively and are determined from the orthogonality condition \( \braket{\psi^+|\psi^+}=\braket{\psi^-|\psi^-}=1 \). Here the energy corresponding to $H^+$ from (\ref{F:H-1D}) is
\begin{align}\label{E+(n,m)}
    E^+_{n,m}=E^+_n-\epsilon_m
\end{align}
From (\ref{F:int1}) and (\ref{F:int2}), the eigenfunctions and energy eigenvalues are related as
\begin{equation}\label{F:psi-1D}
\begin{cases}
    \psi^-_{n+1}(x,m)=\frac{1}{\sqrt{E^+_{n,m}}}\;A^\dagger\psi^+_{n}\quad \text{and }\quad E^-_{n+1,m}=E^+_{n,m};\;E^-_{0,m}=0& \text{for normalizable }\phi_m^{-1}(x)\\
    \psi^-_{n}(x,m)=\frac{1}{\sqrt{E^+_{n,m}}}\;A^\dagger\psi^+_{n}\quad \text{and }\quad E^-_{n,m}=E^+_{n,m}& \text{for non-normalizable }\phi_m(x) \text{ or }\phi_m^{-1}(x)
\end{cases}
\end{equation}

\renewcommand{\theequation}{B.\arabic{equation}} 
\setcounter{equation}{0} 

\section*{Appendix B: SUSY in Higher Dimension}

In higher dimensions, $\frac{d^2}{dx^2}$ is replaced by a Laplacian operator \(\nabla^2 \) and the scalar superpotential $W(x)$ becomes a vector superpotential \( \vec{W} \). We write linear operators in a frame-independent manner \cite{fernandez2004higher, das1997higher} as
\begin{align}\label{F:A-}
A &= \hat{e}^+ \cdot \left(\vec{\nabla} + \vec{W}\right),\qquad  \hat{e}^+=\sum_{i=1}^{D}c_i\;\hat{e}_i\\ \label{F:A+}
A^{\dagger} &= \left(\vec{\nabla}^{\dagger} + \vec{W}^{\dagger}\right)\cdot \hat{e}^{+\;\dagger},\qquad  \hat{e}^{+\;\dagger}=\sum_{i=1}^{D}c_i^{\dagger}\;\hat{e}_i ,
\end{align} 
where \(c_i\) are coefficients which are either 1 or an imaginary number depending on the dimension \(D\) and \( \hat{e}_i \) are orthonormal unit vectors. The form of \(\vec{\nabla}\), \(\mathbf{\hat{e}^+}\) and \(\vec{W}\) in 1D, 2D, and 3D is tabulated in Table \ref{2D-3D}.
\begin{table}[h]
    \centering
    \begin{tabular}{|c|c|c|c|}
    \hline
    \textbf{Dimension}&$\mathbf{\hat{e}^+}$ & $\vec{\nabla}$ & $\mathbf{\vec{W}}$ \\
        \hline
        1D &\( \hat{e}_x \) &\( \hat{e}_x\frac{\partial}{\partial x} \) & \(  \hat{e}_x W_x \) \\
        2D &\( \hat{e}_x + i\hat{e}_y \) &\( \hat{e}_x\frac{\partial}{\partial x} + \hat{e}_y \frac{\partial}{\partial y}\) & \( \hat{e}_x W_x + \hat{e}_y W_y \) \\
        3D &\( i_1\hat{e}_x + i_2\hat{e}_y + i_3\hat{e}_z \) &\(\hat{e}_x\frac{\partial}{\partial x} + \hat{e}_y \frac{\partial}{\partial y}+\hat{e}_z \frac{\partial}{\partial z}\) & \( \hat{e}_x W_x + \hat{e}_y W_y + \hat{e}_z W_z \) \\
        \hline
    \end{tabular}
    \caption{Form of $\hat{e}^+$, $\vec{\nabla}$ and $\mathbf{\vec{W}}$ in higher dimension.}
    \label{2D-3D}
\end{table} 
In 2D, the coefficients $c_i$ are $1$ and $i$ (imaginary number) respectively. In 3D, $c_i$ coefficients follow the quaternionic algebra. Quaternions extend complex numbers as $a+bi_1+ci_2+di_3$. The units $i_k$, where $k$ runs from 1 to 3, and their product satisfies \cite{das1997higher} 
\begin{equation}\label{F:quat-algebra}
    i_j i_k = -\delta_{jk} + \sum_{l=1}^{3}\epsilon_{jkl}\;i_l,
\end{equation}
where \( \epsilon_{jkl} \) is the Levi-Civita symbol. The three $i_k$'s are anti-hermitian
\[i_k^{\dagger}=-i_k\]
The partner potentials when calculated using vector superpotential as defined in (\ref{F:A-}) and (\ref{F:A+}) gives
\begin{align}\label{F:V-W}
    V^{\mp} &= \sum_{k}^{D}W_k^2 \mp \sum_{k}^{D}\frac{\partial W_k}{\partial x_k} + Q_D,
\end{align}
where the superpotentials \( W_k \) along each dimensions are defined in terms of seedless eigenfunction  similar to (\ref{F:W-1D}) as
\begin{equation}\label{F:W}
    W_k = -\left|\frac{\partial}{\partial x_k}\ln[\phi_{m_k}(x_k)]\right|
\end{equation} 
and \( Q_D \) can be constructed for any $D$ dimension, In particular $Q_D$ for $D=1,2$ and $3$ are defined as
\begin{align}\label{F:Q}
    \begin{split}
    Q_1 &= 0\\
    Q_2 &= 2i \left(W_x \frac{\partial}{\partial y} - W_y \frac{\partial}{\partial x}\right)\prod_{k}\phi_{m_k}(x_k) \\
    \text{and}\quad Q_3 &= \mp 2\left[i_1 \left(W_y \frac{\partial}{\partial z} - W_z \frac{\partial}{\partial y}\right) + i_2 \left(W_z \frac{\partial}{\partial x} - W_x \frac{\partial}{\partial z}\right) + i_3 \left(W_x \frac{\partial}{\partial y} - W_y \frac{\partial}{\partial x}\right)\right]\prod_{k}\phi_{m_k}(x_k)
    \end{split}
\end{align}
where \(i_1\), \(i_2\) and \(i_3\) are the Quaternions units. 
The potential in higher dimension $V^+(x,y,\cdots)$ having the form
\begin{equation}\label{F:V+}
    V^+(x_1,x_2,\cdots) = \sum_{k}^{D}V_k^+(x_k)
\end{equation}
where $V_k^+(x_k)$ is the partner potential in each dimension. The eigenfunction $\psi^+_{n_1,n_2,\cdots}(x,y,\cdots)$ will be given by the product of eigenfunctions in each dimension as
\begin{equation}\label{F:psi+}
    \psi^+_{n_1,n_2,\cdots}(x,y,\cdots) = \prod_{k}\psi^+_{n_k}(x_k).
\end{equation}
the energy $E^+_{n_1,n_2,\cdots}$ will be given by
\begin{equation}\label{F:E+}
    E^+_{n_1,n_2,\cdots} = \sum_{k}^{D}E^+_{n_k}\qquad \text{where }n_k\in \mathbb{Z}
\end{equation}
The $m$-dependent SUSY partner potential $V^-(x,y,\cdots,m_1,m_2,\cdots)$ will then be given by
\begin{equation}\label{F:V-}
    V^-(x_1,x_2,\cdots,m_1,m_2,\cdots) = V^+(x_1,x_2,\cdots) - 2\sum_{k}^{D}\left|\frac{\partial^2 \phi_k(x_k)}{\partial x_k^2}\right|+Q_D
\end{equation}
and for vanishing \(Q_D\), the eigenfunction $\psi^-(x_1,x_2,\cdots,m_1,m_2,\cdots)$ will be given by taking the product of eigenfunctions in each dimension as
\begin{equation}\label{F:psi-}
    \psi^-_{n_1,n_2,\cdots}(x_1,x_2,\cdots,m_1,m_2,\cdots) = \prod_{k}\psi^-_{n_k}(x_k,m_k)
\end{equation}
where $\psi^-_{n_k}(x_k,m_k)$ is given by (\ref{F:psi-1D}) along each dimension. Similarly, the energy eigenvalues $E^-(m_1,m_2,\cdots)$ for Hamiltonian $H^-(x,y,\cdots)$ will be given by summing the energy $E^-(m_k)$ corresponding to Hamiltonian $H^-(x_k)$ in each dimension.
\begin{equation}\label{F:E-}
    E^-(m_1,m_2,\cdots) = \sum_{k}^{D}E^-(m_k)
\end{equation}
where \(E^-(m_k)\) is given by (\ref{F:psi-1D}) along each dimension.



\begin{thebibliography}{10}

\bibitem{infeld1951factorization}
Leopold Infeld and TE~Hull,
\newblock The factorization method,
\newblock {\em Reviews of modern Physics}, 23(1):21, 1951.

\bibitem{gendenshtein1983derivation}
L~{\'E} Gendenshte{\^\i}n,
\newblock Derivation of exact spectra of the schrodinger equation by means of
  supersymmetry,
\newblock {\em Jetp Lett}, 38(6):356--359, 1983.

\bibitem{cooper1983aspects}
Fred Cooper and Barry Freedman,
\newblock Aspects of supersymmetric quantum mechanics,
\newblock {\em Annals of Physics}, 146(2):262--288, 1983.

\bibitem{cooper1995supersymmetry}
Fred Cooper, Avinash Khare, and Uday Sukhatme,
\newblock Supersymmetry and quantum mechanics,
\newblock {\em Physics Reports}, 251(5-6):267--385, 1995.

\bibitem{fellows2009factorization}
Jonathan~M Fellows and Robert~A Smith,
\newblock Factorization solution of a family of quantum nonlinear oscillators,
\newblock {\em Journal of Physics A: Mathematical and Theoretical},
  42(33):335303, 2009.

\bibitem{mielnik1984factorization}
Bogdan Mielnik,
\newblock Factorization method and new potentials with the oscillator spectrum,
\newblock {\em Journal of mathematical physics}, 25(12):3387--3389, 1984.

\bibitem{hussin1998simple}
V~Hussin, B~Mielnik, et~al,
\newblock A simple generation of exactly solvable anharmonic oscillators,
\newblock {\em Physics Letters A}, 244(5):309--316, 1998.

\bibitem{gomez2010extension}
David G{\'o}mez-Ullate, Niky Kamran, and Robert Milson,
\newblock An extension of bochners problem: exceptional invariant subspaces,
\newblock {\em Journal of Approximation Theory}, 162(5):987--1006, 2010.

\bibitem{gomez2009extended}
David G{\'o}mez-Ullate, Niky Kamran, and Robert Milson,
\newblock An extended class of orthogonal polynomials defined by a
  sturm--liouville problem,
\newblock {\em Journal of Mathematical Analysis and Applications},
  359(1):352--367, 2009.

\bibitem{quesne2008exceptional}
Christiane Quesne,
\newblock Exceptional orthogonal polynomials, exactly solvable potentials and
  supersymmetry,
\newblock {\em Journal of Physics A: Mathematical and Theoretical},
  41(39):392001, 2008.

\bibitem{odake2009infinitely}
Satoru Odake and Ryu Sasaki,
\newblock Infinitely many shape invariant potentials and new orthogonal
  polynomials,
\newblock {\em Physics Letters B}, 679(4):414--417, 2009.

\bibitem{odake2010another}
Satoru Odake and Ryu Sasaki,
\newblock Another set of infinitely many exceptional laguerre polynomials,
\newblock {\em Physics Letters B}, 684(2-3):173--176, 2010.

\bibitem{odake2013krein}
Satoru Odake and Ryu Sasaki,
\newblock Krein--adler transformations for shape-invariant potentials and
  pseudo virtual states,
\newblock {\em Journal of Physics A: Mathematical and Theoretical},
  46(24):245201, 2013.

\bibitem{yadav2013scattering}
Rajesh~Kumar Yadav, Avinash Khare, and Bhabani~Prasad Mandal,
\newblock The scattering amplitude for a newly found exactly solvable
  potential,
\newblock {\em Annals of Physics}, 331:313--316, 2013.

\bibitem{yadav2015scattering}
Rajesh~Kumar Yadav, Avinash Khare, and Bhabani~Prasad Mandal,
\newblock The scattering amplitude for rationally extended shape invariant
  eckart potentials,
\newblock {\em Physics Letters A}, 379(3):67--70, 2015.

\bibitem{yadav2015group}
Rajesh~Kumar Yadav, Nisha Kumari, Avinash Khare, and Bhabani~Prasad Mandal,
\newblock Group theoretic approach to rationally extended shape invariant
  potentials,
\newblock {\em Annals of Physics}, 359:46--54, 2015.

\bibitem{yadav2016parametric}
Rajesh~Kumar Yadav, Avinash Khare, Bijan Bagchi, Nisha Kumari, and
  Bhabani~Prasad Mandal,
\newblock Parametric symmetries in exactly solvable real and $PT$ symmetric
  complex potentials,
\newblock {\em Journal of Mathematical Physics}, 57(6):062106, 2016.

\bibitem{grandati2012rational}
Yves Grandati,
\newblock Rational extensions of solvable potentials and exceptional orthogonal
  polynomials,
\newblock In {\em Journal of Physics: Conference Series}, volume 343, page
  012041. IOP Publishing, 2012.

\bibitem{marquette2013two}
Ian Marquette and Christiane Quesne,
\newblock Two-step rational extensions of the harmonic oscillator: exceptional
  orthogonal polynomials and ladder operators,
\newblock {\em Journal of Physics A: Mathematical and Theoretical},
  46(15):155201, 2013.

\bibitem{kumar2024rationally}
Rajesh Kumar, Rajesh~Kumar Yadav, and Avinash Khare,
\newblock Rationally extended harmonic oscillator potential, isospectral family
  and the uncertainty relations,
\newblock {\em Annals of Physics}, 463:169623, 2024.

\bibitem{yadava2017rationally}
Rajesh~Kumar Yadav, Nisha Kumari, Avinash Khare, and Bhabani~Prasad Mandal,
\newblock Rationally extended shape invariant potentials in arbitrary $D$-
  dimensions associated with exceptional xm polynomials,
\newblock {\em Acta Polytechnica}, 57(6):477--487, 2017.

\bibitem{clark1984non}
TE~Clark and ST~Love,
\newblock Non-relativistic supersymmetry,
\newblock {\em Nuclear Physics B}, 231(1):91--108, 1984.

\bibitem{de1983supersymmetric}
Marie de~Crombrugghe and Vladimir Rittenberg,
\newblock Supersymmetric quantum mechanics,
\newblock {\em Annals of Physics}, 151(1):99--126, 1983.

\bibitem{andrianov1984factorization}
Alexander~A Andrianov, NV~Borisov, and Mikhail~V Ioffe,
\newblock The factorization method and quantum systems with equivalent energy
  spectra,
\newblock {\em Physics Letters A}, 105(1-2):19--22, 1984.

\bibitem{khare1984supersymmetric}
Avinash Khare and Jnanadeva Maharana,
\newblock Supersymmetric quantum mechanics in one, two and three dimensions.
\newblock {\em Nuclear Physics B}, 244(2):409--420, 1984,

\bibitem{sukumar1985supersymmetry}
CV~Sukumar,
\newblock Supersymmetry and the dirac equation for a central coulomb field,
\newblock {\em Journal of Physics A: Mathematical and General}, 18(12):L697,
  1985.

\bibitem{leblanc1992extended}
Martin Leblanc, G~Lozano, and H~Min,
\newblock Extended superconformal galilean symmetry in chern-simons matter
  systems,
\newblock {\em Annals of Physics}, 219(2):328--348, 1992.

\bibitem{das1997higher}
Ashok Das, S~Okubo, and SA~Pernice,
\newblock Higher-dimensional susy quantum mechanics,
\newblock {\em Modern Physics Letters A}, 12(08):581--588, 1997.

\bibitem{carballo2004polynomial}
Juan~M Carballo, Javier Negro, Luis~M Nieto, et~al,
\newblock Polynomial heisenberg algebras,
\newblock {\em Journal of Physics A: Mathematical and General}, 37(43):10349,
  2004.

\bibitem{morales2015truncated}
VS~Morales-Salgado et~al,
\newblock Truncated harmonic oscillator and painlev{\'e} iv and v equations,
\newblock In {\em Journal of Physics: Conference Series}, volume 624, page
  012017. IOP Publishing, 2015.

\bibitem{louck1973canonical}
JD~Louck, M~Moshinsky, and KB~Wolf,
\newblock Canonical transformations and accidental degeneracy. i. the
  anisotropic oscillator,
\newblock {\em Journal of Mathematical Physics}, 14(6):692--695, 1973.

\bibitem{fernandez2004higher}
David~J Fern{\'a}ndez~C and Nicol{\'a}s Fern{\'a}ndez-Garc{\'\i}a,
\newblock Higher-order supersymmetric quantum mechanics,
\newblock In {\em AIP Conference Proceedings}, volume 744, pages 236--273.
  American Institute of Physics, 2004.

\end{thebibliography}

\end{document}